\documentclass{elsart}

\usepackage{graphicx}

\usepackage{amssymb}

\begin{document}

\begin{frontmatter}

\title{Particle acoustic detection in gravitational wave aluminum resonant antennas}

\author[label1]{B.~Buonomo},
\author[label3,label2,label4]{E.~Coccia},
\author[label2]{S.~D'Antonio},
\author[label1]{G.~Delle~Monache},
\author[label1]{D.~Di~Gioacchino},
\author[label1]{V.~Fafone},
\author[label1]{C.~Ligi},
\author{A.~Marini\thanksref{label1}\corauthref{azm}},
\author[label1]{G.~Mazzitelli},
\author[label1]{G.~Modestino},
\author[label1]{S.~Panella},
\author[label3,label1]{G.~Pizzella},
\author[label1]{L.~Quintieri},
\author[label1]{S.~Roccella{\thanksref{label5}}},
\author[label1]{F.~Ronga},
\author[label1]{P.~Tripodi},
\author[label1]{P.~Valente{\thanksref{label6}}}

\address[label1] {Istituto Nazionale di Fisica Nucleare - Laboratori Nazionali di Frascati, Via~ E.~Fermi,~40 - 00044 Frascati, Italy}
\address[label2] {Istituto Nazionale di Fisica Nucleare - Sezione Roma2, Via~della~Ricerca~Scientifica,~1 - 00133 Rome, Italy}
\address[label3] {Dipartimento di Fisica, Universit\`a di Tor Vergata, Via~della~Ricerca~Scientifica,~1 - 00133 Rome, Italy}
\address[label4] {Istituto Nazionale di Fisica Nucleare - Laboratori Nazionali del Gran Sasso, S.S.~17 ~BIS~km. 18.910 - 67010 Assergi, Italy}

\corauth[azm]{Corresponding author} 
\thanks[label5]{also  Dipartimento di Ingegneria Strutturale e Geotecnica, Universit\`a La Sapienza, Rome, Italy}
\thanks[label6]{now at Istituto Nazionale di Fisica Nucleare - Sezione Roma1, Rome, Italy}

\begin{abstract}
The results on cosmic rays detected by the gravitational antenna NAUTILUS have motivated an experiment (RAP) based on a suspended cylindrical bar, which is made of the same aluminum alloy as NAUTILUS and is exposed to a high energy electron beam. Mechanical vibrations originate from the local thermal expansion caused by warming up due to the energy lost by particles crossing the material. The aim of the experiment is to measure the amplitude of the fundamental longitudinal vibration at different temperatures. We report on the results obtained down to a temperature of about 4 K, which agree at the level of $\sim$10\% with the predictions of the model describing the underlying physical process.
\end{abstract}

\end{frontmatter}

\section{Introduction}
\label{intro}
Cosmic rays generate signals in a massive gravitational wave (GW) detector. The signals are due to vibrations produced by the heating along the particle trajectories, depending on the thermal expansion coefficient and on the specific heat of the material. 

For cylindrical detectors the acoustic model \cite{beron,beron1,milano,cabibbo,malugin,deru,gla,amaldipiz,bar} 
gives quantitative predictions  for the  amplitude of  the longitudinal modes of oscillation due to the energy loss of a particle impinging on the cylinder. The amplitude is  proportional to the ratio of the thermal expansion coefficient to the heat capacity. This ratio is part of the definition of the material Gr\"{u}neisen parameter:
\begin{displaymath}
\gamma=\frac{\beta K_{T} V_{m}}{c_{v}}\ ,
\end{displaymath}

where $\beta$ is the volume expansion coefficient, $K_{T}$ is the bulk module at constant temperature,
$V_{m}$ is the molar volume and $c_{v}$ is the specific heat at constant volume. The parameter $\gamma$ is nearly constant over a wide range of temperatures.

Cosmic ray measurements made with the aluminum GW detector NAUTILUS  gave essentially two types of results:
\begin{itemize}
\item{When the NAUTILUS bar was at a temperature T=0.14 K, that is in a superconductive regime, 
the rate of the high energy signals due to cosmic ray showers was larger by two-four order of magnitude than the expectations based on the models \cite{naut1,naut2} }

\item{When the bar temperature was T=1.5 K, aluminum in a normal state,  few signals were large, most of the signals obeying \cite{naut3} the predictions.}
\end{itemize}

It was not clear to us whether that depended on the behavior of the Gr\"{u}neisen parameter in the superconductive regime (for example, data on the aluminum critical field as a function of pressure and temperature~\cite{harris} suggest  values of the parameter larger than in the normal state~\cite{marini}).
 
To gain information on this problem a new experiment (RAP) was designed, consisting in bombarding a small aluminum bar, whose temperature we could control, with a known electron beam,  provided by the Beam Test Facility (BTF) of the DA$\Phi$NE $\Phi$-factory complex at the INFN Laboratory in Frascati.

We compare the measured maximum amplitude of the bar fundamental longitudinal mode of oscillation with the expected value:

\begin{equation}
B_{th}=B_{0}\ (1+\epsilon)
\label{bth}
\end{equation}
where ~\cite{milano}:
\begin{equation}
B_{0}=\frac{2}{\pi}\frac{\alpha L}{c_v M} W
\label{b0}
\end{equation}

Here $\alpha$ is the linear thermal expansion coefficient ($\beta=3\alpha$ for aluminum), $L$,  $M$ are respectively  length,  mass of the cylinder and $W$ is the total energy released by the beam to the bar. The term $\epsilon$ accounts for corrections estimated by Monte Carlo methods due to contributions $O[(R/L)^{2}]$, $R$ being the cylinder radius,  and to the beam structure (see Appendix~\ref{montecarlo}).

After a short description of the experimental setup and a discussion on the  beam simulated interactions within the bar, we report  on the results obtained down to a temperature of $\sim$ 4 K. 

\section{Experimental setup}
\label{setup}

The RAP setup is composed by the beam~\cite{mazbtf}   and the detector~\cite{5ea,frontier}.  The cylindrical test mass, the suspension system, the cryogenic system, the transducer and the data acquisition system are parts  of the detector.

\subsection{Beam}
\label{beam}

\begin{figure}[htbp]
\begin{center}
\includegraphics[width=0.40\linewidth]{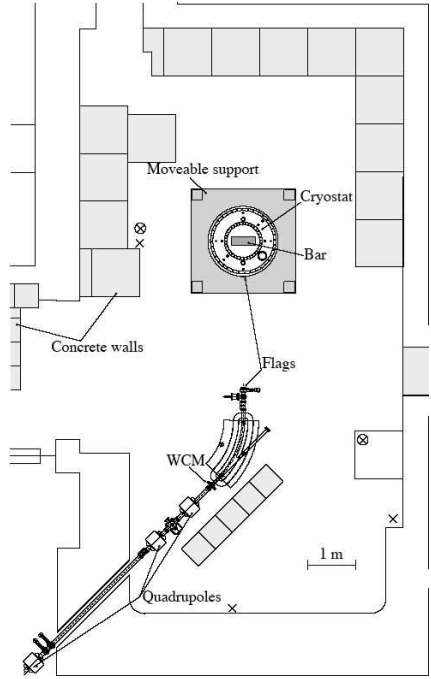}
        \caption{\it  Schematic layout of the RAP setup in the BTF hall.}
        \label{fig1}
\end{center}
\end{figure}

DA$\Phi$NE BTF provides the controlled beam for RAP. Although the facility is optimized for the production of single electrons (or positrons), mainly for high energy detector calibration and testing purposes, it can provide particles in a wide range of energy (25-750 MeV) and intensity (up to $\sim 10^{10}$ particles/pulse).   Particles can be produced in pulses of 1 or 10 ns duration, with a fairly uniform distribution. The maximum energy of the beam is $\sim$ 510 MeV when BTF is  jointly operated  with the injections into the DA$\Phi$NE collider.
The installation site of RAP is 2.5 m far away from the exit of the beam (see Fig.~\ref{fig1}). 
The position of the beam at the exit of the line and at the entrance of the cryostat was monitored shot-by-shot by two high sensitivity fluorescence flags, 1 mm thick alumina doped with chromium targets. The spot size at the entrance of the cryostat, 50 cm far from the center of the bar, is $\sim$ 2 cm  in diameter.
The Monte Carlo simulation of the detector (see Section \ref{simula}) indicates that the beam spot at the surface of the cylindrical bar mantains almost this dimension, due to negligible  effects induced by the thickness of the cryostat vacuum shields intercepted by the beam. 

BTF delivers to RAP single pulses containing $O[10{^8}]$ electrons each. In this range standard beam diagnostics devices to monitor the beam charge have low sensitivity, while saturation affects particle detectors.  Therefore a monitor (BCM),  based on an integrating current transformer readout by a charge digitizer,  was installed  to measure the pulse charge. The device accuracy is $\sim$3\% and the sensitivity, dominated by the digitizer, is $\rm{\sim1.4\cdot 10{^7}}$ electrons, corresponding to a $\sigma\sim 4\cdot 10{^6}$ electrons. The device is equipped with a calibration coil used to control the gain, the noise and the time shaping of the generated signals.

\subsection{Test mass and suspension system}
\label{tmss}
 The oscillating test mass is  a  cylindrical bar (diameter 0.182 m, length 0.5 m, mass 34.1 kg) made of Al5056, the same aluminum alloy (5.2 Mg\%, 0.1\%Mn, 0.1\% C) used for NAUTILUS. The resonance frequency of the first longitudinal mode of vibration is 5096 Hz at T=296 K.

The suspension  is an axial-symmetric system of seven attenuation stages in series made of copper.
The system  provides  a -150 dB attenuation of the external mechanical noise in a frequency window spanning from 1700 to 6000 Hz. The symmetry axis of the suspension passes through the center of mass of the cylinder. The suspension is linked to the lateral surface of the cylinder by means of two brass screws at the mid-section of the cylinder.

\subsection{Cryogenic system}
\label{cryo}

The cryogenic system is based on a commercial cylindrical cryostat (3.2 m height, 1.016 m diameter) and a $^3HeÐ^4He$ dilution refrigerator (base temperature=100 mK, cooling power at 120 mK=1 mW), to be installed in the near future. 
Cryogenic operations start  by the filling of the cryostat with liquid nitrogen, both in the nitrogen and in the helium dewars, followed by the cool-down to 4.2 K with liquid helium.
The thermal exchange among liquid helium and the bar is realized by filling the space surrounding the bar with $\sim$ 1 mbar of gaseous helium. This is indeed the only thermal link between the bar and the liquid helium, the suspension-bar assembly being mechanically disconnected from the cryostat, with the exception of three thin stainless steel wires connecting the two components.

Temperatures are measured  by two platinum resistor and three ruthenium dioxide resistor thermometers, in the 60-300 K range and in the 0-60 K range respectively, with a precision of 0.1~K.
All the thermometers are controlled by a multi-channel resistance bridge.

A mechanical structure encloses the cryostat allowing an easy positioning of the detector on the beam line and the consequent removal after the expiration of the dedicated periods of data taking.

\begin{figure}[htbp]
\begin{center}
\includegraphics[width=0.40\linewidth]{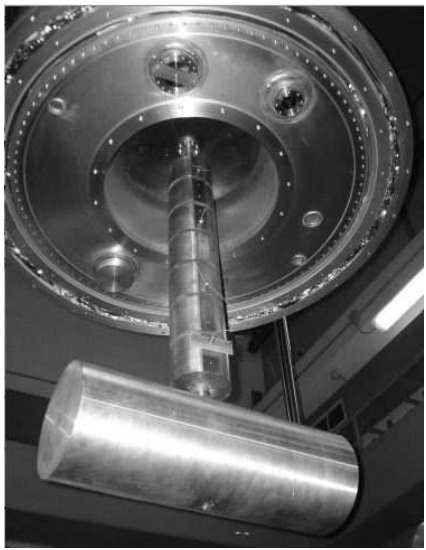}
        \caption{\it  The suspended bar in the opened cryostat.}
        \label{fig2}
\end{center}
\end{figure}

Fig.~\ref{fig2} shows the bar suspended inside the upper half of the cryostat.

\subsection{Transducer and data acquisition system}
\label{pz}

Two Pz24 piezoelectric ceramics (0.01x0.01x0.005 m$^3$), electrically connected in parallel,  are inserted in a slot cut in the position opposite to the bar suspension point and are squeezed when the bar contracts.  In this configuration the stress measured at the bar center  is related to the displacement of the bar faces. The generated voltage signal is sent to an amplifier ($V_{noise}\sim $4 nV$/\sqrt{\rm Hz}$).  The data acquisition system (DAQ), based on a peak sensing 16-bit VME ADC and a VME Pentium III CPU running Linux, collects and stores on disk amplified data coming from the transducer and data originated from the thermometers and from the beam monitoring system. 
The acquisition system provides also tools for performing a fast inline analysis of the collected data.

\section{Calibrations}
\label{cal}
\begin{figure}[htbp]
\begin{center}
\includegraphics[width=0.6\linewidth]{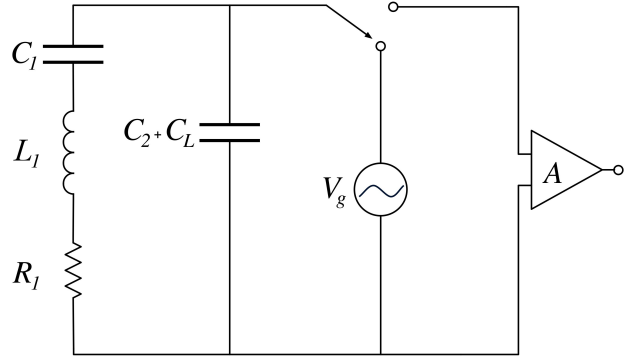}
        \caption{\it  Equivalent electrical circuit and calibration scheme.}
        \label{fig3}
\end{center}
\end{figure}

It is possible to measure the value of $\lambda$, the transducer electro-mechanical coupling factor, making use of the electric equivalent circuit shown in Fig.~\ref{fig3}. 

This circuit includes a $R_1- L_1- C_1$ series that expresses the bar characteristics (dissipation, mass, elasticity) and in parallel the Pz24's with capacity $C_2$. To measure the equivalent circuit parameters we proceed as follows:
\begin{itemize}
\item{We apply a signal $V_g~cos(\omega_o~t)$ at the Pz24's for a time $\Delta t$ much smaller than the decay time of the oscillations, where $\omega_o=2\pi f_o={1}/{\sqrt{L_1C_1}}$ is the angular resonance frequency of the bar.}

\item{We switch the Pz24's from the oscillator producing $V_g$ to the  amplifier and measure the signal $V_o~cos(\omega_o~t+\phi)$ due to the bar oscillation.}
\end{itemize}

It is possible to show~\cite{equi}  that $C_1$ is given by the following equation:

\begin{displaymath}
C_1=\frac{V_o}{V_g \Delta t \pi f_o}[C_2+C_L]
\label{cuno}
\end{displaymath}

\noindent where  $C_L$ is the load capacity that includes the cable and the input impedance of the amplifier. Having measured the value of $C_1$ we obtain $\lambda$ for the first mode of longitudinal oscillation by:

\begin{equation}
\lambda=\frac{2 \pi f_o}{C_2+C_L} \sqrt{\frac{M C_1}{2}}
\label{lambda}
\end{equation}
 
 \noindent where M is the mass of the bar.

The values of $\lambda$, obtained at different temperatures and before the collection of data useful for the estimation of the vibrational effects induced by the beam, are shown in Table~\ref{tb1} together with the measured frequency  of the first mode of longitudinal oscillation $f_o$ and the its decay time $\tau_o$.

\begin{table}
\begin{center}
\begin{tabular}{|c|c|c|c|}
\hline
 $T[K]$& $f_o[Hz]$  &$ \tau_o [s]$ &  $\lambda[10^{7}V/m]$ \\
\hline
264           &5143.7  & 6.25 & 1.32\\
71             &5397.3  & 24 & 1.48 \\
4.5            &5412.7  & 84  & 1.32 \\
\hline

\end{tabular}
\caption{\it Frequency, decay time of the first mode of longitudinal oscillation and transducer coupling factor as a function of temperature.
}
\label{tb1}
\end{center}
\end{table}

The PZ24's are inserted at a distance of $\sim$ 0.09 m from the cylinder axis, while the calibration procedure we have used is correct in the $(R/L)^{2}\ll1$ approximation. We checked the calibration procedure inserting a commercial calibrated accelerometer  with 5\% accuracy  at the center of one of  the bar end-surfaces.  At room temperature the displacement measured by the PZ24's was 2.4\% smaller than the one measured by the accelerometer and at $T\sim$~77 K the displacement was $\sim$6\% smaller.
Consequently we assume a 6\% systematic error  in measuring the displacement amplitudes.

\section{Simulation of the beam energy loss in the aluminum bar}
\label{simula}

\begin{figure}[htbp]
\begin{center}
\includegraphics[width=0.6\linewidth]{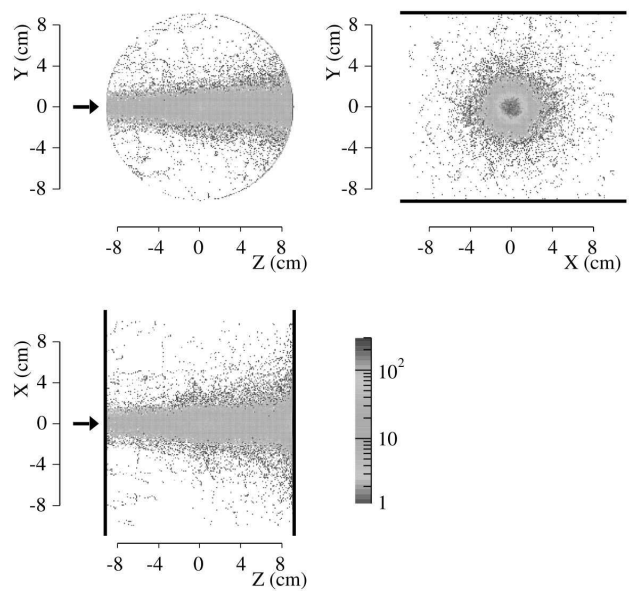}
        \caption{\it  Monte Carlo simulation: secondary
	particles distribution for 510 MeV primary electrons impinging
        on the bar for the three projections. The black arrow
	indicates the direction of the incoming beam.}
        \label{fig4}
\end{center}
\end{figure}

We have developed a Monte Carlo simulation using the CERN package GEANT 3.21, taking into account the real geometry and material of the cryostat and of the bar, and a realistic parametrization of the BTF beam, i.e. the beam spot size and
divergence, and simulating the passage of electrons through the matter. 
The size of the beam spot and its angular and momentum spread have been simulated according to the measured characteristics: the well focussed BTF beam has a fairly Gaussian shape in the transverse plane with $\sigma_x = \sigma_y = 0.005$ m, the momentum spread is $\sim$ 1\% and the beam emittance is $\epsilon= 1$ mm mrad. The cryostat has been schematized as a succession of three coaxial aluminum cylinders.  Due to the presence of these aluminum shields, with a total Al depth of
$1.7 \cdot 10^{-2}$~m, the energy of the electrons is slightly degraded by the emission of Bremsstralhung photons.  On average, more than 100 secondary particles are produced for each 510 MeV primary electron entering the bar.
The spatial distribution of these secondary particles inside the bar is shown in Fig.~\ref{fig4} for the three perpendicular projections. The development of the electromagnetic shower is clearly visible; indeed the bar diameter corresponds to $\sim$ 2 radiation lengths.

The total energy released to the bar  by an electron in the beam is then evaluated adding the contribution of the energy lost  by all the $N_i$ secondary particles ($j=0$ is the contribution of the primary electron itself), at each of the $M_j$ tracking steps in the simulation code:

\begin{displaymath}
\Delta E= \sum_{j=0}^{N_i}\sum_{k=1}^{M_j}\Delta E_{jk}
\end{displaymath}

\begin{figure}[htbp]
\begin{center}
\includegraphics[width=0.6\linewidth]{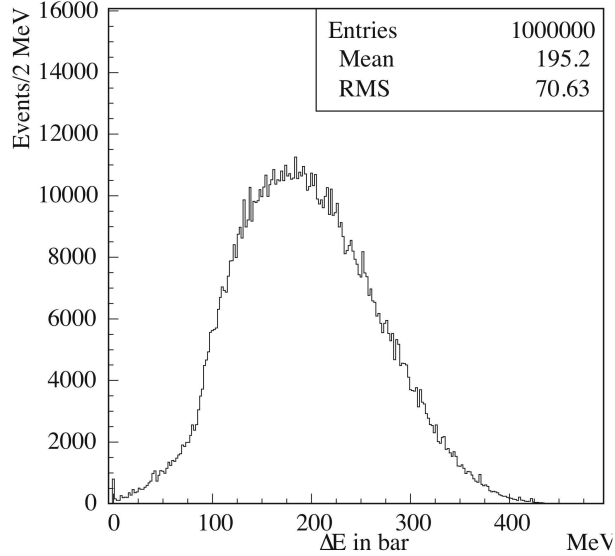}
        \caption{\it  Monte Carlo simulation for 510 MeV primary electrons impinging
        on the bar:  $\Delta E$ per electron }
        \label{fig5}
\end{center}
\end{figure}

The distribution of the energy lost  by one electron with energy of 510~MeV in 10$^6$ trials, as shown in Fig.~\ref{fig5}, leads to an average energy loss per electron:

\begin{equation}
<{\Delta  E}> \pm \  \sigma_{\Delta E}=195.2 \pm 70.6 \ MeV
\label{wsing}
\end{equation}

and for a beam composed by N electrons to a total energy loss:

\begin{equation}
  W=N <{\Delta  E}> \  , \ \sigma_{W}=\sqrt{N}\  \sigma_{\Delta E}
  \label{wtot}
\end{equation}

We performed extensive simulations (see Appendix~\ref{montecarlo}) in order to: (a)~apply the acoustic model to each infinitesimal step of the tracked particles, (b)~include the treatment of the longitudinal modes of oscillation at $O[(R/L)^{2}]$, (c)~evaluate the effects due to the beam structure. This study leads to a value of -0.04 for $\epsilon$ in equation~(\ref{bth}).

\section{Measurements and comparison with expectations}
\label{meas}
Several electron pulses were applied to the bar at various temperatures (264~K, 71~K, 4.5~K) with DAQ collecting ADC data, thermometer data and number of electrons per pulse.
The maximum amplitude of the first longitudinal mode of oscillation of the bar, $B_{m}$, is related to experimental measured values according to the relation: 

\begin{displaymath}
B_{m}=V_{max}/(G \lambda)
\end{displaymath}

where $V_{max}$ is the maximum amplitude of the component of the signal at the frequency corresponding to the first longitudinal mode, obtained using Fast Fourier Transform procedures applied to the ADC stored data,  $G$  is the gain of the amplifier and $\lambda$ is the coupling factor defined in equation~(\ref{lambda}).

Knowledge  of $\alpha(T)$ and $c_{v}(T)$  for aluminum is needed for computing the relation (\ref{b0}) at different temperatures: spline interpolations on data of ref.~\cite{kroeg} (12 K$<T\leq$300~K) and the parametrization in ref.~\cite{coll} ($T\leq$12 K)  give $\alpha(T)$, while $c_{v}(T)$ is obtained by spline interpolations on values of $c_{p}$ reported in ref.~\cite{crc}. Table~\ref{tb2}  shows the calculated values of $\alpha$,  $c_{v}$, $B_{0}$ due to a deposition of 10$^{-3}$~J in a RAP-like bar made of  pure aluminum.

\begin{table}
\begin{center}
\begin{tabular}{|c|c|c|c|}
\hline
 $T[K]$& $\alpha [10^{-6} K^{-1}]$ & $c_{v}\ [J\  mol^{-1} K^{-1}]$  & $B_{0} [10^{-13} m] $\\
\hline
264           &22.2               & 23.5                        &  2.32 \\
71             &7.5                 &7.94                         &  2.32 \\
4.5            &5.8\ 10$^{-3} $& 7.6\ 10$^{-3}$   & 1.88  \\
\hline

\end{tabular}
\caption{\it $B_{0}$ due to W=10$^{-3}$ J and input values for the calculation in case of pure aluminum.}

\label{tb2}
\end{center}
\end{table}

\begin{figure}[htbp]
\begin{center}
\includegraphics[width=0.60\linewidth]{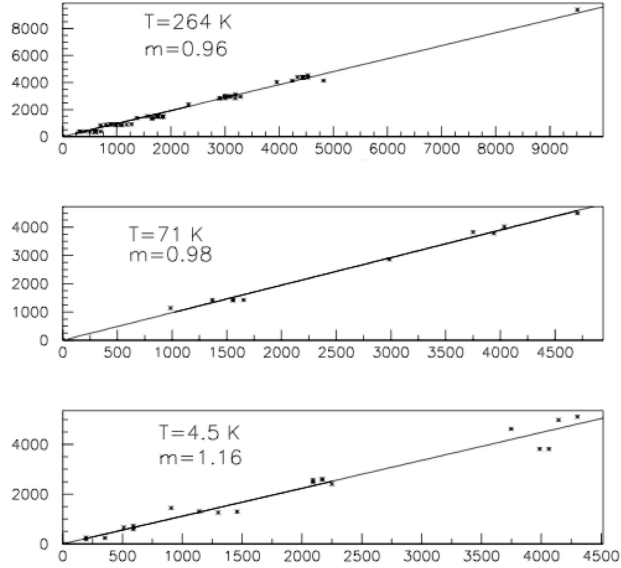}
        \caption{\it Measured maximum amplitudes [10$^{-16}$m] of induced vibrations, $B_{m}$ (vertical), versus the expected ones, $B_{th}$ (horizontal), for various beam intensities. Straight lines are the least square fits with equation $B_{m}=m\ B_{th}$.}
        \label{fig6}
\end{center}
\end{figure}

The correlations among measured $B_{m}$ and estimated $B_{th}$ (equation(\ref{bth}))  are shown in Fig.~\ref{fig6} and the values of the parameter $m$ fitting the relation $B_{m}=m\ B_{th}$ are shown in Table~\ref{tb3}. A systematic error of $\sim$~7\%, obtained from the quadrature of the beam monitor (3\%) and $\lambda$ determination (6\%) accuracies, affects the measurements.
\begin{table}
\begin{center}
\begin{tabular}{|c|c|c|}
\hline
 $T[K]$& $m$ & $\Delta m$  \\
\hline
264           &0.96       & 0.01  \\
71             &0.98       & 0.03  \\
4.5            &1.16       & 0.03 \\
\hline

\end{tabular}
\caption{\it Values of m fitting $B_{m}=m\ B_{th}$.}
\label{tb3}
\end{center}
\end{table}

\section{Conclusions}
\label{conc}
The results we obtained for the maximum amplitude of the fundamental mode of longitudinal vibration of the bar are in good agreement with the expectations coming from the thermo-acoustic model describing the particle energy loss conversion into mechanical energy in the temperature range 270-4 K. This is the first time that experimental results on thermo-acoustic energy conversion in Al5056 are obtained below 270 K using our technique. The result at $T\sim 270$~K improves the agreement with the model when compared to the findings of  past experiments \cite{milano,holland} based on the same technique. Amplitude measured values at $T\sim 4$~K are slightly higher than expectations: this is probably due to a different behavior  of Al5056 respect to pure aluminum or to an imperfect knowledge of the input parameters for the calculations in a region of temperatures where the ratio $\alpha/c_{v}$ sensibly depends on $T$.

We are greatly indebted to Dr.~S.~Bertolucci for the continuos support given to the experiment, to Drs. D.~Babusci and G.~Giordano for having derived the formal expression
of the divergences used in the simulations of Appendix~\ref{montecarlo} and to Prof.~M.~Bassan for many useful discussions.
We would like to thank Messrs.~G.~Ceccarelli, R.~Ceccarelli, M.~De~Giorgi for the help given during the cryogenic operations of the detector and Messrs.~M.~Iannarelli and ~R.~Lenci which helped with  the experiment readout setup. 


\appendix

\section{Monte Carlo simulation} 
\label{montecarlo}
The energy in the normal modes of order $n$ of a thin cylindrical bar 
(radius $R$ much smaller than the length $L$) of mass $M$ 
excited by a ionizing particle, can be calculated using 
the thermo-acoustic model~\cite{bar} in terms of
the sound source $\Sigma$, of the thermo-mechanical parameters of the
bar and a geometrical factor $G_n$:
\begin{equation}
E_n=\frac{1}{2M} \left( \frac{L \cdot G_n}{v_s} \cdot \Sigma \right)^2
=\frac{1}{2M} \left( \frac{L \cdot G_n}{v_s} \cdot \gamma \frac{dE}{dx}\right)^2
\label{energy}
\end{equation}
where $\gamma$ is the Gr\"uneisen parameter, 
$v_s$ is the speed of sound, $dE/dx$ is the specific energy loss.
The geometrical factor $G_n$, taking into account the path $\Lambda$
inside the bar and the coupling to the eigen-modes of frequency
$\omega_n$, is defined as the path integral
of the divergence of the displacement vector (normalized to the volume) 
$\vec{u}_n$:
\begin{equation}
G_n=\frac{v_s}{\omega_n}\frac{1}{\Lambda}\int_{\Lambda}
\left(\vec{\nabla}\cdot\vec{u}_n\right)dl
\label{gn}
\end{equation}
In other terms, the energy for the $n$-th mode is obtained by squaring the amplitude of the
vibration for a path of a single ionizing particle inside the bar:
\begin{displaymath}	
E_n = \frac{1}{2M}\frac{\gamma^2}{\omega_n^2} \cdot A^2_n
\end{displaymath}
The corresponding expression when the bar is crossed by $N$ particles 
is given by summing (incoherently) over all the amplitudes,
\begin{displaymath} 
A_n=\sum_{i=1}^{N} A^{(i)}_n\mbox{,}
\end{displaymath}
so that Eq.~(\ref{energy}) can be written as:
\begin{equation}
E_n=\frac{1}{2M} \frac{\gamma^2}{\omega^2_n} 
\left( \sum_{i=1}^N A^{(i)}_n \right)^2
\label{energyi}
\end{equation}
In the estimate of the expected amplitude when a beam of
particles crosses the bar, two extreme approaches are possible: 
the specific energy loss is assumed to
be constant all along the trajectory inside the resonant bar,
so that it can be computed separately from $G_n$;
or the real energy loss is evaluated for each infinitesimal
step and is multiplied by the correct geometrical factor
before summing all the amplitudes:
\begin{displaymath}	
A_n=\sum_{i=1}^N \int_{{\Lambda}_i} \left(\vec{\nabla}\cdot\vec{u}_n\right)
\frac{dE}{dl_i}dl
\end{displaymath}	
The latter method requires a
simulation of the elementary processes of energy loss 
inside the material.
The GEANT routines are able to track all the
secondary particles produced in the interaction with the material
of the generated (primary) particle, and 
all the different physical processes occuring when the particles cross
the matter are simulated, taking into account the appropriate cross-sections
for the different materials in the set-up, and the current energy and
momentum of the particle: ionization, Coulomb scattering, 
and Bremsstrahlung for charged particles; pair production, Compton scattering, 
and photoelectric effect for photons. 

All the secondary particles produced in the interactions with the
matter are followed until their energy falls below a given
threshold. The size of the step of the simulation for each particle is 
automatically adjusted, in order to optimize the accuracy vs. the computing
time, taking into account the set-up geometry and materials together with the
kinematical quantities of the currently tracked particle.

Such a simulation code is well suited to include the 
calculation of the geometrical integral in the tracking routines, 
together with the evaluation of the specific energy loss for the different 
physical processes happening inside the materials, at each step of the 
simulation. 
Moreover, since the contribution of each single particle in the beam 
has to be added in an incoherent sum, it is easy to extend the sum over 
all the secondary particles of the electromagnetic shower that develops 
inside the bar for each incident electron.

The total excitation amplitude for the $i$-th primary electron in the beam 
is then evaluated adding the contribution of all the $N_i$ secondary 
particles ($j=0$ is the contribution of the primary electron itself), 
at each of the $M_j$ tracking steps in the simulation code:
\begin{displaymath}
A_n^{(i)}=\sum_{j=0}^{N_i}\sum_{k=1}^{M_j} a_n^{(jk)}
\end{displaymath}
By summing together the contribution of all the steps of all the tracked particles, 
quantities such as the total path length:
\begin{displaymath}
\Delta x= \sum_{j=0}^{N_i}\sum_{k=1}^{M_j}\Delta \vec{x}_{jk}\mbox{,}
\end{displaymath}
and the total energy loss: 
\begin{displaymath}
\Delta E= \sum_{j=0}^{N_i}\sum_{k=1}^{M_j}\Delta E_{jk}\mbox{,}
\end{displaymath}
can be calculated
for each incoming primary electron. 

The integral in the expression of $G_n$ is numerically calculated for 
each step, assuming that the divergence of the displacement vector 
is constant (this is a good approximation for steps of $\sim$ 1~mm):
\begin{equation}
a_n^{(jk)} = \int_{{\Lambda}_j}\left(\vec{\nabla}\cdot\vec{u}_n\right)\frac{dE}{dl_j}dl_j =
\sum_{k=1}^{M_j} \left(\vec{\nabla}\cdot\vec{u}_n(\vec{x}_{jk})\right) \cdot \Delta E_{jk}
\mbox{.}
\label{ajk}
\end{equation}

In order to compute the divergence $\vec{\nabla}\cdot\vec{u}_n(\vec{x}_{jk})$, the
exact solution of the normal modes of a finite cylindrical bar have been replaced
by the approximate solution for an infinite cylinder due to Pochhammer and Chree (PC). This 
approximation holds for a thin cylinder, i.e. as long as
$\alpha_n=n\pi \frac{R}{L} < 1$.
The PC solution has been expanded to the order ${O}[\alpha_n^2]$ leading to:
\begin{displaymath}
\vec{\nabla}\cdot\vec{u}_n(\vec{x}_{jk})=
\frac{\sqrt{2}\alpha_n}{R}\left[1+\frac{\sigma(1-\sigma)}{4}\alpha_n^2\right]\left[2\sigma-1+\frac{\sigma}{2}\left(\alpha_n\frac{r}{R}\right)^2\right] sin\left(\alpha_n\frac{z}{R}\right)
\end{displaymath}

where $\sigma$ (=0.347 for isotropic aluminum at room temperature) is the Poisson module and $r,z$ are referred to a cylindrical coordinate system having the $z$-axis coincident with the bar axis and the $z$-origin located at one of the bar end-surfaces.

In order to simulate a number of beam pulses on the RAP bar, the vibrational amplitudes
in the different normal modes have been computed simulating a given number of electrons
in the beam, with the nominal energy of 510 MeV and with the proper beam spread and divergence. 
This gives an estimate of the bar response for a given beam intensity. The
following results refer to 100 pulses of $10^4$ electrons/pulse, hitting the bar at
the center. 
\begin{figure}[htbp]
\begin{center}
\includegraphics[width=0.90\linewidth]{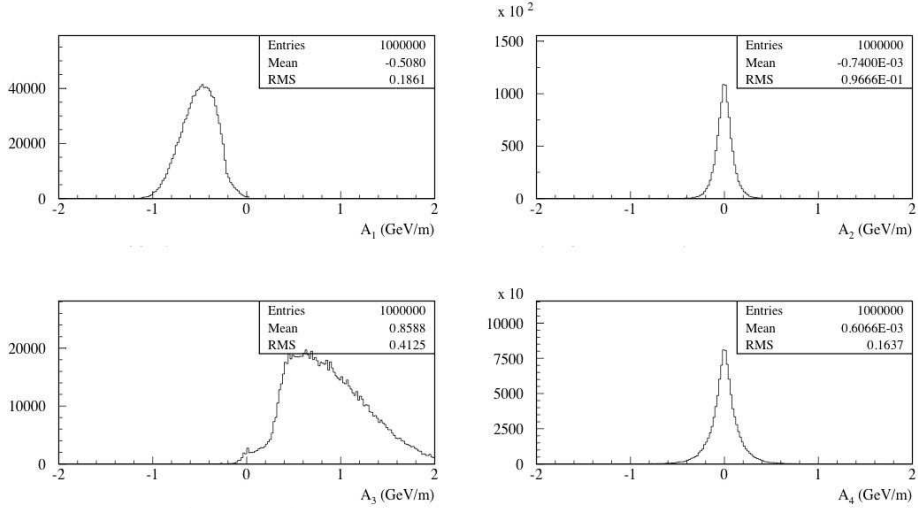}
        \caption{\it Monte Carlo results for the 
	first four modes of the amplitude: on the left the odd modes ($n=1,3$), 
        on the right the even modes ($n=2,4$) distributions (on top, bottom respectively).
	In the histograms all the events from 100 simulated shots of $10^4$ electrons with
	$E=510$ MeV
	have been accumulated.}
        \label{fig:a_i}
 \end{center}
\end{figure}
The distributions of the $A_n$ amplitudes for the first four modes are shown 
in Fig.~\ref{fig:a_i}.
The odd modes have a non-zero average, while the the average for the
even modes is consistent with zero. This is what we expect from the properties
of the eigen-functions, that are symmetric (anti-symmetric) with respect
to $z=L/2$ for odd (even) modes. Since the beam enters the bar at $z=L/2$, $A_{n=1,3,\ldots}$ will  
have the same sign; while $A_{n=2,4,\ldots}$ will change sign and have
a null expectation value.

In order to obtain an estimate of  $\epsilon$ (equation~(\ref{bth})) from this full simulation we compare the average value of $A_1$=--0.508~GeV/m,  derived from the corresponding histogram of Fig.~\ref{fig:a_i},
against the value of  $A_1$ when computed $O[\alpha^1]$:
\begin{displaymath}
A_1=\frac{\sqrt{2}\pi}{L}(2\sigma-1)<\Delta E> = -0.530\ \rm{GeV/m}\ ,
\end{displaymath}

which is related to equation~(\ref{b0}) in the hypothesis that all the energy lost by an electron (cfr.
equation~(\ref{wsing})) is deposited at the bar center. Thus we obtain $\epsilon\sim$ -0.04.

\end{document}